# Optimizing Spectral Phase Transfer in Four-Wave Mixing with Gas-filled Capillaries

HAO ZHANG[1,2,*], LIN-SHAN SUN[1], JACK HIRSCHMAN[2,3], MIRALI SEYED SHARIATDOUST[1,6], FEDERICO BELLI[4], SERGIO CARBAJO[1,2,5,6]

[1]*Department of Electrical & Computer Engineering, University of California Los Angeles, Los Angeles, CA 90095, USA*
[2]*SLAC National Accelerator Laboratory, Stanford University, Menlo Park, California 94025, USA*
[3]*Department of Applied Physics, Stanford University, CA 94305, US*
[4]*School of Engineering and Physical Sciences, Heriot-Watt University, Edinburgh EH14 4AS, Scotland, United Kingdom*
[5]*Physics and Astronomy Department, University of California Los Angeles, CA 90095, USA*
[6]*California NanoSystems Institute, Los Angeles, CA 90095, USA*

*\*haozh@g.ucla.edu*

**Abstract:** Four-wave mixing (FWM) in gas-filled hollow-core capillaries, a nonlinear optical process that mixes signal and pump photon frequencies to generate idler frequency photons, offers a method for precise spectral phase transfer from signal to idler at ultrashort timescales and extreme powers. However, this regime is challenged by competing linear and nonlinear dynamics, leading to significant trade-offs between spectral phase transfer and conversion efficiency. Our computational investigation focuses on the upconversion of femtosecond pulses from the infrared (IR) to the ultraviolet (UV), a range notoriously difficult to manipulate. We explore an intermediate energy regime that strikes an optimal balance between FWM-mediated phase-transfer fidelity and nonlinear conversion efficiency. By adjusting the energy ratios and spectral phase profiles of the input signal, we achieve conversion efficiencies of approximately 5-15% while maintaining an effective quasi-linear spectral phase transfer. These findings contribute to establishing first-principles and scaling laws essential for applications such as high-precision imaging, spectroscopy, quantum transduction, and distributed entangled interconnects, facilitating advanced control of ultrafast photonic and electronic wavepackets in quantum materials with unprecedented spatial and temporal precision.

## 1. Introduction

Nonlinear optical effects and ultrafast optical physics are pivotal in both fundamental scientific research and emerging technologies in medicine and communications, providing insights into the interactions between light and matter[1–6] and enabling new states of light. These fields are intricately connected to the development of advanced optical systems, where the precise control of spectral phase is a key factor[1,7]. Spectral phase control is not only crucial for ultrafast pulse shaping[7–10] and high-resolution spectroscopy[11,12] but also plays a significant role in optical computing[13,14] and quantum control[15,16]. Currently, most of the spectral phase control experiments are performed in the visible (VIS) and near-IR (NIR) wavelengths[17,18] by using the phase control devices, such as spatial light modulator (SLM)[19,20] or acoustic-optics modulators [21,22]. Experimental validation of spectral phase control has historically been challenging due to the large reduction in overall device efficiency, particularly affecting specific wavelength regimes such as the UV. Moreover, these devices struggle at high peak powers or average powers due to their low damage thresholds. Therefore, they are typically used either pre-amplifier, which limits their applicability, or post-amplifier, which constrains the achievable power output[9,23–25]. While some studies have demonstrated that plasmonic metasurfaces[26] and second harmonic nanoparticles[27] can control phase in the UV range, the intrinsic damage threshold and challenging phase matching conditions also pose limitations to advancements in this field[18,28]. Moreover, several applications require a highly modulated spectral phase (e.g. a sine-modulated phase for electron shaping[29]), which is not achievable by using conventional simple dispersive media, such as prism pairs or gratings[30]. To summarize, the principal challenges in implementing spectral phase control encompass limitations related to peak and average power handling capabilities, the difficulty of establishing complex phase relationships, and restrictions on encoding specific shapes for optimization, high-speed transformations, or artificial intelligence and machine learning-based methodologies[31,32]. These limitations collectively hinder the broader applicability and effectiveness of current spectral phase control techniques in advanced optical systems[23–25].

An alternative approach that circumvents these limitations relies on an indirect spectral phase control method, focusing on transferring a desired spectral phase from one spectral region to another[8,18]. This transfer is effectively achieved through frequency mixing, where a pulse with a precision-controlled spectral phase is combined with another pulse of narrow bandwidth. Integrating this spectral phase transfer methodology with frequency conversion and amplification techniques holds particular promise, as it can ideally optimize the balance between conversion efficiency and phase-matching bandwidth. A quintessential method of this integration is the application of phase-matched four-wave mixing (FWM) in hollow core capillary fibers (HCC)[33–35]and anti-resonant fibers [36,37] filled with noble gas. Gas-filled fibers are an ideal platform for frequency conversion and spectral phase manipulation. Their features include tunable dispersion properties, a high damage threshold, and tunable output wavelengths, all underpinned by a strong nonlinear response[36,38,39]. These attributes collectively make them well-suited for optical manipulations, aligning with the demands of cutting-edge photonics research[8,18,38].

While significant progress has been made in the realm of FWM within gas-filled HCC[33,34], there remains an essential challenge: apprehending the trade-offs between spectral phase transfer and efficiency in these systems. This study addresses this critical issue, bridging the gap between theoretical insights and practical implementations in spectral phase engineering. To enhance performance and expand the utility of this technology in diverse scientific domains, including high-precision spectroscopy[40] and advanced communication technologies[1,41], our investigation explores two distinct operational energy ratio regimes to highlight the trade-off between spectral phase transfer and conversion efficiency. In both regimes, the pump spectral distribution remains unaltered and only the signal is modified, in which we apply different phase modulations: linear, quadratic, sinusoidally modulated phase and π-step phase. Note that we consider only shaping of the spectral phase, without amplitude modulation of the spectrum. For each spectral phase shape, we identify operation regimes in which (1) large conversion efficiencies are maximized; (2) signal-to-idler linear spectral phase mapping is enhanced; and (3) the emergence of a narrow region where the two above are balanced, that is, quasi-linear phase transfer from the input signal to the output idler is achieved at moderate conversion efficiencies. Our work aims to offer a detailed analysis of these regimes and establish a benchmark reference to enhance the effectiveness of spectral phase control using FWM in the HCC systems, an essential guideline for precise spectral phase manipulation in areas like inverse optical engineering and ultrafast pulse shaping.

2. **Methods**

**Degenerate FWM model.** We consider the seeded and degenerate FWM case, which means two equal pump photons are involved, and one signal photon is seeding the FWM process in the gas-filled HCC. We can describe the nonlinear effect via the nonlinear Schrödinger equation (NLSEs)[42]:

$$\frac{\partial A_p}{\partial z} + \underbrace{\frac{i}{2}\beta_{2p}\frac{\partial^2 A_p}{\partial T^2}}_{\text{SOD}} - \underbrace{\frac{1}{6}\beta_{3p}\frac{\partial^3 A_p}{\partial T^3}}_{\text{TOD}} + \underbrace{\frac{1}{2}\alpha_p A_p}_{\text{Loss}} = i\gamma_p \left( \underbrace{(|A_p|^2 + 2(|A_s|^2 + |A_i|^2))A_p}_{\text{SPM} \quad \text{XPM}} + \underbrace{2A_s A_i A_p^* \exp(i\delta\beta z)}_{\text{FWM}} \right)$$
(1)

$$\frac{\partial A_s}{\partial z} + d_{sp}\frac{\partial A_s}{\partial T} + \underbrace{\frac{i}{2}\beta_{2s}\frac{\partial^2 A_s}{\partial T^2}}_{\text{SOD}} - \underbrace{\frac{1}{6}\beta_{3s}\frac{\partial^3 A_s}{\partial T^3}}_{\text{TOD}} + \underbrace{\frac{1}{2}\alpha_s A_s}_{\text{Loss}} = i\gamma_s \left( \underbrace{(|A_s|^2 + 2(|A_p|^2 + |A_i|^2))A_s}_{\text{SPM} \quad \text{XPM}} + \underbrace{A_p^2 A_i^* \exp(-i\delta\beta z)}_{\text{FWM}} \right)$$
(2)

$$\frac{\partial A_i}{\partial z} + d_{ip}\frac{\partial A_i}{\partial T} + \underbrace{\frac{i}{2}\beta_{2i}\frac{\partial^2 A_i}{\partial T^2}}_{\text{SOD}} - \underbrace{\frac{1}{6}\beta_{3i}\frac{\partial^3 A_i}{\partial T^3}}_{\text{TOD}} + \underbrace{\frac{1}{2}\alpha_i A_i}_{\text{Loss}} = i\gamma_i \left( \underbrace{(|A_i|^2 + 2(|A_p|^2 + |A_s|^2))A_i}_{\text{SPM} \quad \text{XPM}} + \underbrace{A_p^2 A_s^* \exp(-i\delta\beta z)}_{\text{FWM}} \right)$$
(3)

where, $A_p$, $A_s$ and $A_i$ are the complex amplitudes of pump, signal and idler pulse; $T$ is the time measured in the reference frame moving with the pump pulse; $\alpha$ is the confinement loss of hollow capillary fiber, calculated at pump, signal and idler band; $d_{sp}$ and $d_{ip}$ are the temporal walk-off parameter, which stands for the group velocity mismatch between the pump and signal or the idler. $\beta_2$ and $\beta_3$ are the 2$^{nd}$ and 3$^{rd}$ dispersion coefficients; $z = [0, L]$, where L is the propagation length; $\delta\beta$ is the phase mismatch term and $\gamma$ is the nonlinear parameter, which can be obtained by $\gamma_j = n_{nl}\omega_j/(cA_{eff})$. $n_{nl}$ is the nonlinear index coefficient, $A_{eff}$ is the effective mode area of the fiber. The first term on the right-hand side of the equation represents self-phase modulation (SPM). The second and third terms correspond to cross-phase modulation (XPM) and FWM, respectively.

**Spectral Phase transfer via FWM.** By considering the slowly varying envelope approximation[42], the complex amplitudes of the pump, signal, and idler can be calculated by the NLSEs Eq(1-3). Using those equations, the electrical field of the idler pulse $E_i$ can be approximately described by:

$$E_i \propto i E_p^2 \cdot E_s^\dagger$$
(4)

where $E_s^\dagger$ is the complex conjugate of the signal's electrical field. Building on this relationship, we consider a scenario where the pump wave is relatively monochromatic ($\Delta\lambda_p \setminus \lambda_p \approx 4 \times \Delta\lambda_s \setminus \lambda_s$), characterized by a much longer pump pulse duration compared to the signal pulse (continue wave approximation). Under this assumption, the phase of the idler wave, $\varphi_i(\omega_i)$, can be deduced from the signal's spectral phase. We expand the spectral phase of the signal pulse by Taylor expansion:

$$\varphi_s(\omega) = \varphi_s(\omega_s) + (\omega - \omega_s)\varphi'_s(\omega_s) + \frac{1}{2}(\omega - \omega_s)^2 \varphi''_s(\omega_s) + \frac{1}{6}(\omega - \omega_s)^3 \varphi'''_s(\omega_s) + \ldots \tag{5}$$

From equation (4), the phase of the generated idler pulse after FWM can be written as:

$$\begin{aligned}\varphi_i(\omega) = \varphi_i(\omega + \delta\omega) &= \tfrac{\pi}{2} + 2\varphi_p(\omega_p) - \varphi_s(2\omega_p - (\omega_i + \delta\omega)) \\ &= \underbrace{\tfrac{\pi}{2} + 2\varphi_p(\omega_p)}_{\text{Constant}} - \varphi_s(\omega_s - \delta\omega)\end{aligned} \tag{6}$$

where $\omega_i = 2\omega_p - \omega_s$ and $\delta\omega$ is a small expansion around the idler carrier. Combining equations (5) and (6), it becomes evident that components of the signal described by even functions are transferred with opposite signs, while those represented by odd functions are transferred with the same sign[8,18]. The transferred spectral phase is essentially determined by the bandwidth of the pump spectrum, a consequence of its convolution with the pump spectrum itself.

**Simulation framework.** Our simulation framework is based on the NLSE implemented in Python to model nonlinear optical phenomena. The refractive index values required for these simulations were calculated using the Sellmeier equations[43,44]. In our simulations, we varied the gas pressure within the fiber from 0 to 1 bar, set the fiber radius at 50 μm and conducted the experiments at room temperature (293.15 K), to reflect usual laboratory conditions. We here use a typical core radius as in the previous implementation[36]. The simulation involved computing the dynamics of pulses as they traverse the fiber. For this purpose, we employed a dual-method approach, integrating both the Split Step Fourier Method (SSFM)[38] and the Fourth-Order Runge-Kutta (RK4) method[44].

**Calculation of phase match condition (Ar-only).** A critical aspect of our study is to ensure effective phase matching in the FWM process. To achieve this, we utilized the Nelder-Mead simplex algorithm[45], valuable for its adaptive parameters. This choice facilitated a detailed and fast optimization process, which was crucial for pinpointing the specific conditions that minimize phase mismatch in our simulations. Further, to provide a comprehensive understanding, we have included in the supplementary information (SI) an analysis of how phase mismatch varies under different pressures, temperatures, and fiber radii using Argon (Ar) as the nonlinear mixing medium.

**Input Energy Ratio Regimes.** In our study, we designed the experiment in two distinct energy ratio regimes: high energy ratio and low energy ratio. We set the signal pulse energy at 10 μJ as a constant under the best possible phase mismatch conditions (see Figure S1 in SI). Here, for the convenience, we define the pump-to-signal pulse energy ratio as follows:

$$\rho = \frac{\text{Pump pulse energy}}{\text{Signal pulse energy}} \tag{7}$$

In the high energy regime, we consider $\rho \geqslant 10$ and in the lower energy ratio regime, we consider $\rho \leqslant 1$.

**Conversion efficiency.** One key aspect of our study is the investigation of the detailed balance between conversion efficiency and spectral phase transfer linearity in FWM. To facilitate a consistent and comparative analysis, we maintained a constant energy for the signal pulse across all simulations. This approach provides a stable baseline for evaluating the conversion efficiency, which we quantify as the ratio of the idler pulse energy to the signal pulse energy:

$$\nu = \frac{\text{Idler pulse energy}}{\text{Signal pulse energy}} \tag{8}$$

This simple yet effective metric allows for a clear assessment of efficiency in energy conversion.

**Evaluating the linearity of signal-to-idler phase transfer.** The most critical element in our study is the analysis of spectral phase information transfer, for which we utilized unwrapped phase data. Unwrapped phase data effectively circumvents the ambiguities typically associated with phase wrapping, thereby ensuring an unambiguous and continuous representation of phase evolution throughout the process. We also calculate the Pearson correlation coefficient[46] for a quantitative assessment of phase linearity at the first-order spectral phase transfer in SI, focusing on the $\Delta \lambda$ of the pulse spectrum. The coefficient can be calculated as:

$$P = \frac{\sum (X_i - \bar{X})(Y_i - \bar{Y})}{\sqrt{\sum (X_i - \bar{X})^2 \cdot \sum (Y_i - \bar{Y})^2}} \tag{9}$$

where $X_i$ and $Y_i$ are individual data points of the idler phase and input signal phase respectively. $\bar{X}$ and $\bar{Y}$ are the means of arrays respectively.

For the quadratic, sinusoidally modulated phase, we compare their spectral phase shape qualitatively only because defining a correlation coefficient is difficult.

## 3. Results

Here, we consider the degenerate FWM case, assuming collinearly polarized pump (SHG) and signal (IR) waves at angular frequencies $\omega_p$ and $\omega_s$. When the pump and the signal pulse are launched into the HCC, the idler pulse is generated at the angular frequency $\omega_i = 2\omega_p - \omega_s$, as shown in Figure 1a (input: left; output: right).

To optimize conversion efficiency, our setup is designed to allow tuning of parameters such as gas type, pressure, temperature, and fiber length (fixed after the fabrication) to achieve phase-matching conditions. In our study, we use pump and signal pulse wavelengths of 515 nm ($\Delta \lambda_p$: 2 nm) and 1030 nm ($\Delta \lambda_s$: 12 nm) with 220 fs pulse duration, respectively, resulting in an output idler wavelength of 343 nm ($\Delta \lambda_i$: 4.5 nm). The signal pulse energy is consistently maintained at 10 µJ and pump pulse energy will vary between 10-100 µJ. At room temperature (T = 293 K), phase matching is attained at approximately 0.63 bar when the fibers are filled with Ar (calculations for other noble gases are provided in the SI).

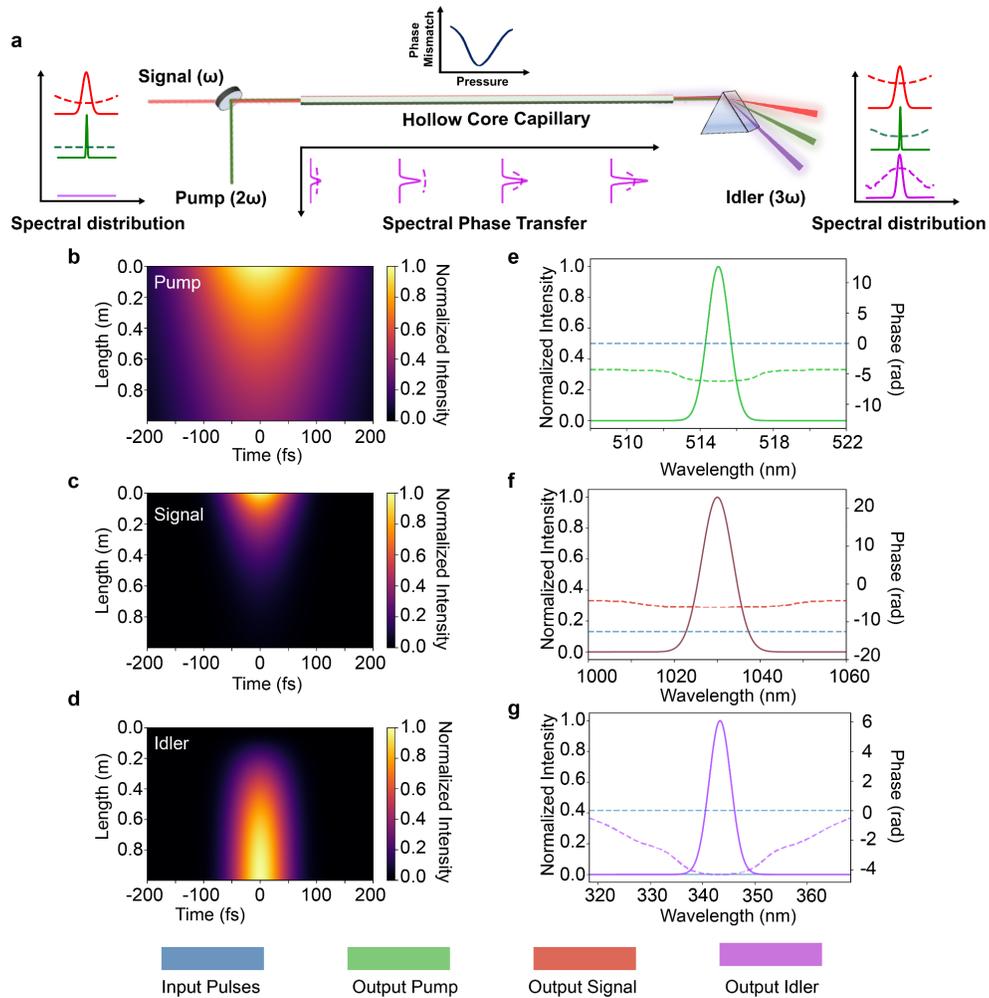

Figure 1. (a) Schematic of FWM in a gas-filled HCC. The optimal phase match condition (inset) can be achieved with Ar at room temperature, approximately at 0.63 bar. (b-d) Temporal intensity evolution of pump, signal and idler pulses (x-axis) as they propagate along (y-axis) the Ar-filled hollow core capillary. (e-g) Comparative analysis of the input and output spectra, along with the phase of the pump, signal, and idler pulses, respectively.

The typical FWM configuration with a flat and zero phases at the input for both the pump and signal pulses, serves as the baseline description [8,17,18] and the trivial evolution of temporal pulse intensity along the HCC for the pump, signal, and idler pulses are shown in Figure 1 (c-e). Notably, the signal pulse depletes rapidly due to a high loss coefficient ($\alpha \simeq 3.67$), due to the waveguide loss. Despite the signal loss, the energy transfer and upconversion to the idler take place. Indeed, while the signal pulse diminishes quickly, the idler pulse energy reaches its maximum at the end of the fiber (approximately $L = 1m$). The corresponding input/output spectral amplitudes and phases are illustrated in Figure 1 (f-h). Due to the nonlinear process (SPM and XPM), the pump phase acquires a parabolic shape at the output. Thus, to investigate the trade-off between phase transfer and conversion efficiency, we will explore a high- and low-energy pump regime and three distinct phase cases involving linear, quadratic, and sinusoidally modulated phases. Additionally, we elaborate further on π-step phase transfer in the SI.

## Case Study 1: linear spectral phase transfer

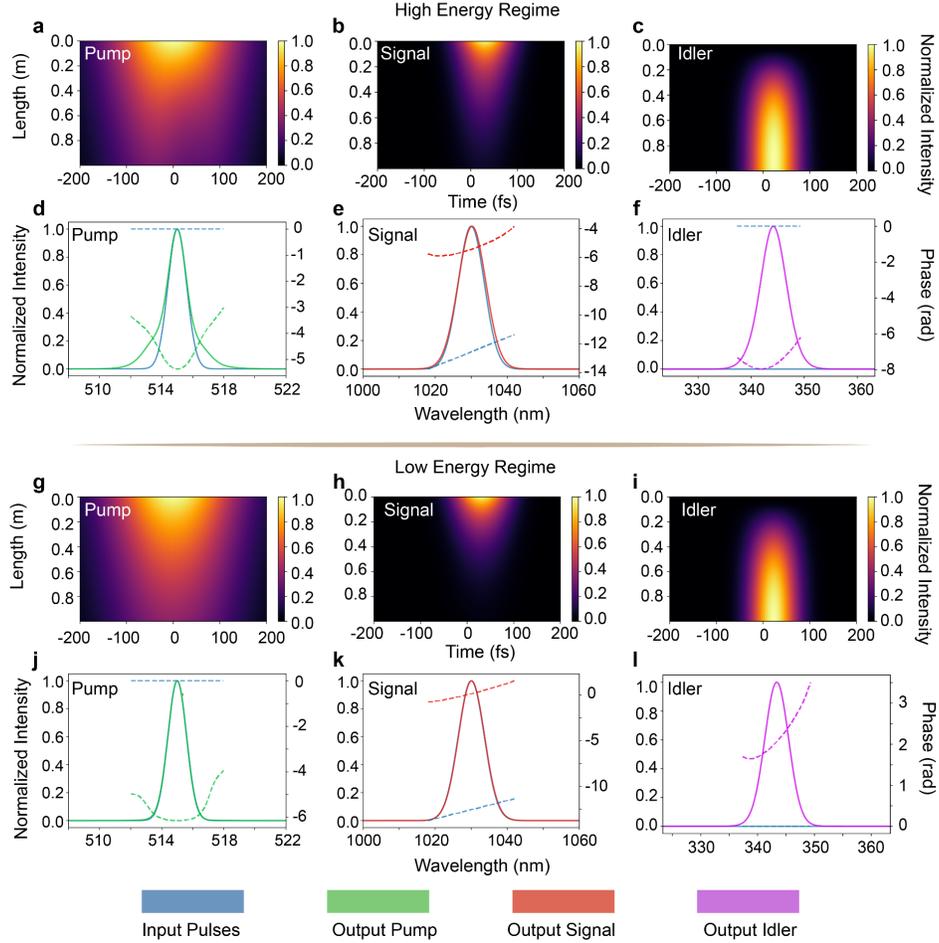

Figure 2. An examination of two distinct scenarios involving spectral and phase comparisons in a gas-filled HCC with the signal pulse energy consistently maintained at 10 μJ. In the first scenario (a-f), a larger conversion efficiency is realized by letting $\rho = 10$. Regardless of the input signal phase's opening direction, the idler phase adopts a parabolic shape due to SPM and XPM. The second scenario (g-l) focuses on enhancing the spectral phase transfer from the signal to the idler through a pulse energy ratio $\rho = 1$. This demonstrates the successful transfer of the signal phase to the idler, including the monotonically increasing spectral phase The $\nu$ recorded for the first and second scenarios are 42.1% and 0.8%, respectively.

In this study, we apply a linear phase to the input signal pulse while maintaining a zero phase for the input pump. We observe that the quality of the transferred phase is influenced by the input energy levels of the input pulses. Nonlinear effects such as SPM and XPM lead to the broadening of the pump and signal spectra, as shown in Figures 2d and e. The idler's spectral phase takes on a parabolic shape, indicating that the linear phase from the signal pulse is not perfectly transferred to the idler. In this high-energy regime, the conversion efficiency $\nu$ is 42.1%. Conversely, in the low energy regime, for $\rho = 1$, we still generate an idler pulse at the fiber's output, as shown in Figure 2g-i. Since the pump energy is much lower than in the high-energy case, the nonlinear effects and accumulated nonlinear phase on the pump remain small, as shown in Figure 2j-k. This regime yields a conversion efficiency $\nu$ of around 0.8%, significantly lower compared to the previous high-energy regime. In this case, the signal phase is transferred directly from the signal to the idler, as evidenced in Figure 2l. Following the theory of phase transfer, the sign of the transferred phase remains consistent with that of the input signal phase.

**Case Study 2: quadratic spectral phase transfer**

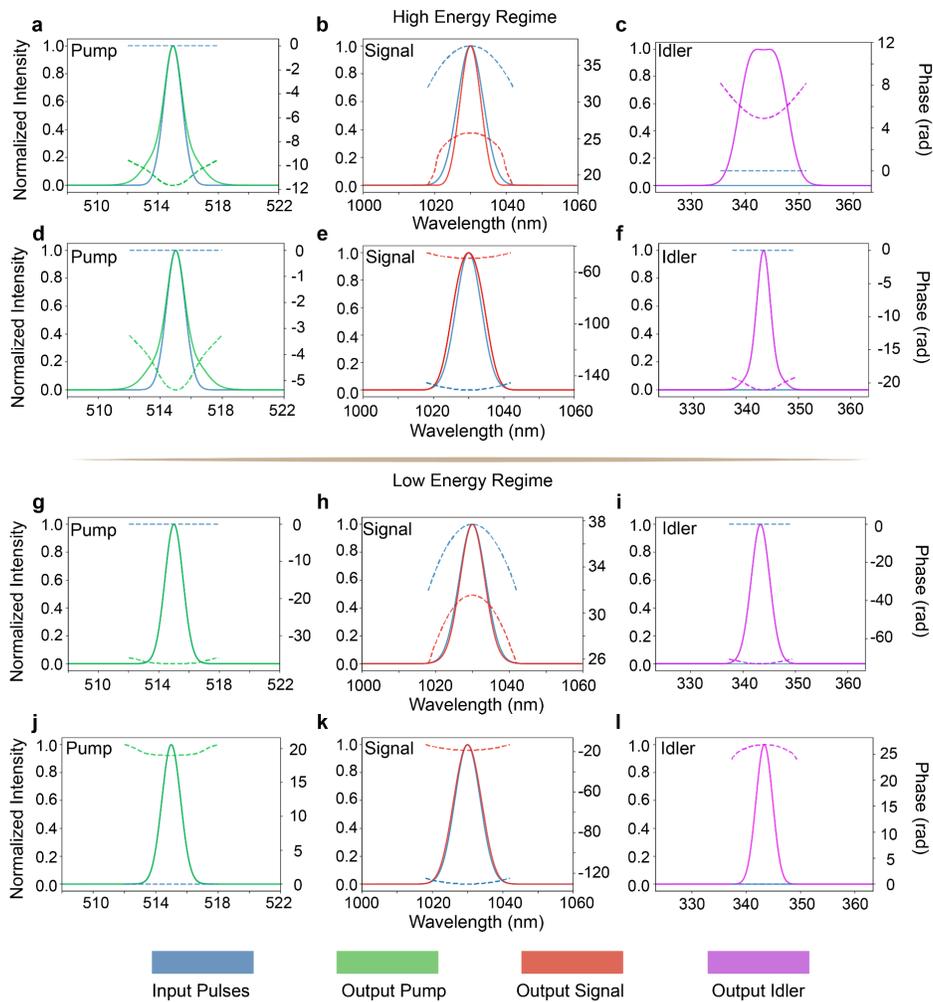

Figure 3. Comparative analysis of the input and output spectra and phases for pump, signal, and idler pulses under two distinct operational regimes. (a-f) Detail the scenario achieving high conversion efficiency with $\rho = 10$, while (g-l) focus on a regime with enhanced spectral phase transfer from signal to idler with $\rho = 1$. The signal pulse energy is consistently set at 10 µJ. Specifically, (a-c) and (g-i) illustrate cases where the initial phase of the signal is a downward-opening parabola (negative coefficient), and (d-f) and (j-l) demonstrate cases with an upward-opening parabolic initial phase of the signal (positive coefficient).

When exploring the transfer of a quadratic phase from the signal to the idler, as depicted in Figure 3, we also categorize the process into two distinct energy ratio regimes: Figures 3a-f represent the high energy regime, while Figures 3g-l illustrate the low energy regime. Specifically, in Figures 3a-c and d-f, we apply two contrasting types of quadratic phase to the signal pulse: one with a downward-opening parabola (up-chirped) and the other with an upward-opening parabola (down-chirped). In the high energy regime, as observed in Figures 4a and d, there is a noticeable broadening of the pump spectra.

As previously mentioned in the Method section, phases described by even functions are expected to be transferred with the opposite sign. However, due to the influence of nonlinear effects, the transferred phases in the high energy regime exhibited a similar parabolic shape (nonlinear effect dominated), not displaying the expected opposite-sign behavior. This phenomenon is evident in Figure 3c and f. In contrast, when operating in the low energy regime, where nonlinearity during propagation is reduced, the quadratic spectral phase transfers more effectively from the signal input to the idler output. As we expected, we observed that a downward-opening phase in the signal translates to an upward-opening phase in the idler and vice versa. This result aligns with our theoretical expectations regarding spectral phase transfer. Regarding conversion efficiency, there is a stark difference between the two regimes. In the high-energy regime, the conversion efficiency $\nu$ stands at 33.2%, while in the low-energy regime, $\nu$ drops significantly to 0.4%.

**Case Study 3: the sinusoidally modulated linear spectral phase transfer**

In our study, when the signal pulse has a sinusoidally modulated linear phase, which is crucial for coherent control[47,48], the high-energy regime (Figure 3e, k) still produces an idler pulse at the fiber's end. This is evident in the pulse evolution images shown in Figure 4a-c, illustrating the effect of the FWM process. Both the pump and signal spectra undergo broadening due to nonlinear effects, clearly depicted in Figures 4d and e. The idler's spectral phase shows a parabolic shape, aligning with our previous discussions, but it also retains some sinusoidal characteristics. In this scenario, the targeted phase does not effectively transfer from the signal to the idler (Figure 4f) as in the previous case and the conversion efficiency $\nu$ is recorded at about 26.3%.

In contrast, in the low energy regime, the sinusoidally modulated linear spectral phase is transferred from the signal to the idler with high fidelity, as shown in Figures 4k-l. Figure 4g-i demonstrates that the output energy of the idler reaches its maximum at the fiber's output, achieving a conversion efficiency $\nu$ of 0.30%. This contrast between the high and low $\rho$ underscores the critical role of energy levels in determining the efficiency and quality of spectral phase transfer during the FWM process.

In this scenario, after the phase modulation, we notice that the signal spectra peaks have a redshift. When $\rho$ becomes larger, the signal spectra split into three peaks, which is due to the interference and coupling effects caused by phase differences introduced by the sinusoidal phase modulation and nonlinear effects. In the time domain, the pulse was split into three sub-pulses, which can be attributed to the interference effect introduced by sinusoidal phase modulation. In the low-energy case, the idler pulse is modulated by the sinusoidal phase. In the high-energy case, although the idler has a quadratic phase, it is also clear that the pulse is modulated by a small sinusoidal phase.

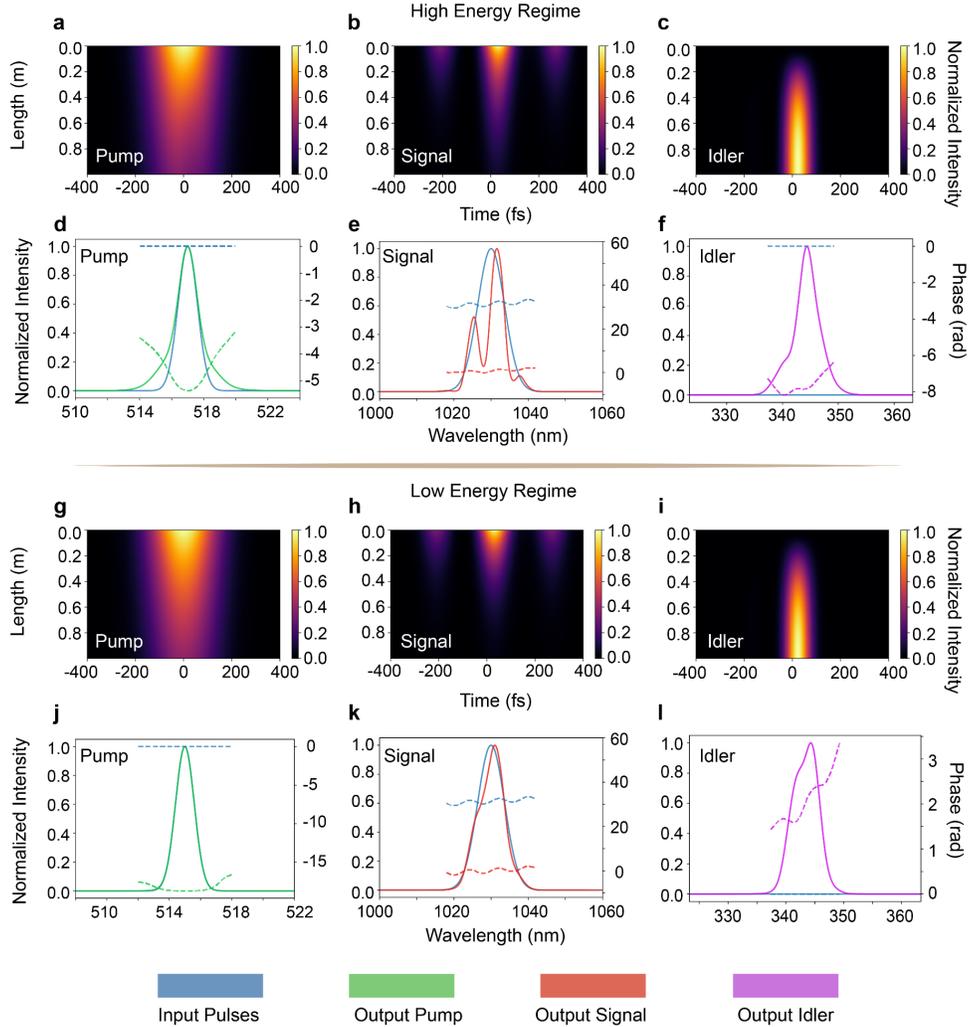

Figure 4. Comparative Analysis of Pulse Evolution and Spectral Characteristics in a Gas-filled Hollow Core Capillary with a sinusoidally modulated linear phase, and a consistent signal pulse energy of 10 μJ Case 1 (Panels a-f): This case focuses on achieving higher conversion efficiency with $\rho = 10$. Panels (a-c) illustrate the evolution of the pump, signal, and idler pulses along the capillary. Panels (d-f) provide a detailed comparison of the input and output spectra and phases for these pulses. In this scenario, the observed conversion efficiency $\nu$ is 26.3%. Case 2 (Panels g-l): Here, the emphasis is on enhancing spectral phase transfer from signal to idler with $\rho = 1$. Panels (g-i) show the progression of the pump, signal, and idler pulses through the capillary. Panels (j-l) present a comparative analysis of the input and output spectra and phases under this condition. The $\nu$ achieved in this case is 0.30%.

## 4. Discussion

When investigating the FWM in gas-filled HCCs, it is crucial to control the input power and frequency precisely in order to achieve the best operational conditions in terms of efficient phase matching or improved conversion efficiency. Derived from Equation (4), it is evident that the idler pulse is proportional to the pump's intensity. Once the SPM and XPM effects are dominant and the pump is no longer monochromatic, that means the transferred phase will be strongly influenced when filtered by the convolution with the pump's spectrum. The nonlinear induced phase[42] will drive equation (4) to:

$$E_i \propto i||E_p| \cdot exp[i\phi_{nonlinear}]|^2 \cdot E_s^\dagger \qquad (9)$$

$$\phi_{nonlinear} \propto \gamma_p P_{pump} L + 2\gamma_s P_{signal} L \qquad (10)$$

where L is the propagation distance along the fiber and $P_{pump/signal}$ is the peak power of the pump/signal pulse. In scenarios where the pump's energy remains lower level, the nonlinear phase will be minimal, thereby favoring the spectral phase transfer from the signal to the idler. Conversely, the increase in the pump's energy lead to a stronger nonlinear phase shift of the pump and on the signal, which directly alters the mapping of the input signal phase onto the idler ones. We here summarize the three cases in Table 1:

**Table 1. Different spectral phase transfer under two energy ratio regimes**

|  | Case-1 linear Signal pulse | | Case-2 quadric signal pulse | | Case-3 sine modulated signal pulse | |
| --- | --- | --- | --- | --- | --- | --- |
| Input energy ratio ($\rho$) | 1 | 10 | 1 | 10 | 1 | 10 |
| Conversion efficiency ($\nu$) | 0.8% | 42.1% | 0.4% | 33.2% | 0.3% | 26.3% |
| Transferred spectral phase shape | Linear | parabolic | quadric with opposite chirp | parabolic | sine | parabolic |

*Signal pulse energy is all set at 10 μJ.

The input pulse energy is crucial in determining the transition between operational regimes. At low energy levels, the system prioritizes direct phase transfer, which is beneficial for precise phase alignment. Thus, the balance between phase transfer and conversion efficiency in FWM is fundamentally tied to the input energy levels. This provides a theoretical guide for optimizing experiments to achieve specific application goals. Especially, in the Case 1 study, we fix the signal pulse energy values to a constant value, respectively to analyze the linearity of the transferred spectral phase of the idler spectrum across various $1 \leqslant \rho \leqslant 10$ shown in Figure S2. We seek to find a trade-off between conversion efficiency and the quality of the transferred phase ($P$). The results indicate that as the $\rho$ increases, the quality of the transferred phase diminishes. We find that when the input energies of the signal and pump are comparable, the transferred spectral phase retains high quality, but at the expense of conversion efficiency. Conversely, higher pump energy relative to the signal pulse enhances light conversion efficiency but compromises the spectral phase transfer quality. This delineates a clear trade-off between spectral phase transfer quality and light conversion efficiency during the FWM process in the HCC. Then we can achieve conversion efficiencies of approximately 5-15% while maintaining an effective quasi-linear spectral phase transfer (Figure S2). We also explore the impact of the on the pump spectrum, as detailed in the SI (Figure S3). It appears that the quality of the transferred phase is inversely proportional to the $\Delta\lambda_p$, while the conversion efficiency directly correlates with it. As shown in Figure S3, when the pump energy is less than the signal energy, the ideal linear spectral phase transfer point will appear at the narrowest pump spectral width. Our study further explores π-step spectral phase transfer, as illustrated in Figure S4 in the SI. Additionally, frequency conversion is achievable by selecting appropriate seed frequencies. Compared to crystals, gases offer a higher damage threshold and a more flexible tuning mechanism, which is advantageous for high-energy ultrafast applications.

5. **Conclusion**

We investigated the quality of spectral phase transfer and high conversion efficiency and have found an inherent trade-off among the two controlled by input energy. We considered three different scenarios: linear spectral phase transfer, quadratic spectral phase transfer, and sinusoidally modulated linear spectral phase transfer. However, it was observed that at higher pulse energies, the idler pulse's quality of spectral phase transfer is reduced when the up-conversion process is maximized. On the other hand, if the main aim is to transfer the phase information from the signal to new idler pulses, then the conversion efficiency would be substantially reduced for a direct quasi-linear transfer. From the simulation results, the highest recordable efficiencies are 26%-43% obtained at a high level of energy with reduced spectral phase transfer contribution. At a lower energy level, the conversion efficiencies decrease down to 0.3%-0.8% with corresponding phase modifications providing optimal phase transfer. Therefore, a balance must be sought when both transferred phase quality and conversion efficiency are critical. This paper shows an example of how to choose a proper setting to either focus on conversion efficiency or improve spectral phase control. It lays the groundwork for upcoming experiments and uses in ultrafast optics.

**Acknowledgment.** The author thanks the support from UCLA and SLAC National Accelerator Laboratory, the U.S. Department of Energy (DOE), the Office of Science, Office of Basic Energy Sciences. The author also extends gratitude to Wenwen Zhang for engaging in comprehensive discussions on the theory section and coding of the simulation framework.

**Funding.** The U.S. Department of Energy (DOE), the Office of Science, Office of Basic Energy Sciences under Contract No. DE-AC02-76SF00515, No. DE-SC0022559, No. DE-FOA-0002859, the National Science Foundation under Contract No. 2231334. F.B. acknowledges support from the Royal Academy of Engineering through Research Fellowship No. RF/202021/20/310 and from the Royal Society (RGS\R1\221085).

**Supporting Information.** See Supplement for supporting content.

**Declarations.** The authors declare no competing interests.

**Data availability.** Data underlying the results presented in this paper are not publicly available at this time but may be obtained from the authors upon reasonable request.

# Supporting information: Optimizing Phase Transfer in Four-Wave Mixing with Gas-filled Capillaries: A Trade-off Study

**S-I: Phase mismatch variations under different pressures and temperatures using Ar**

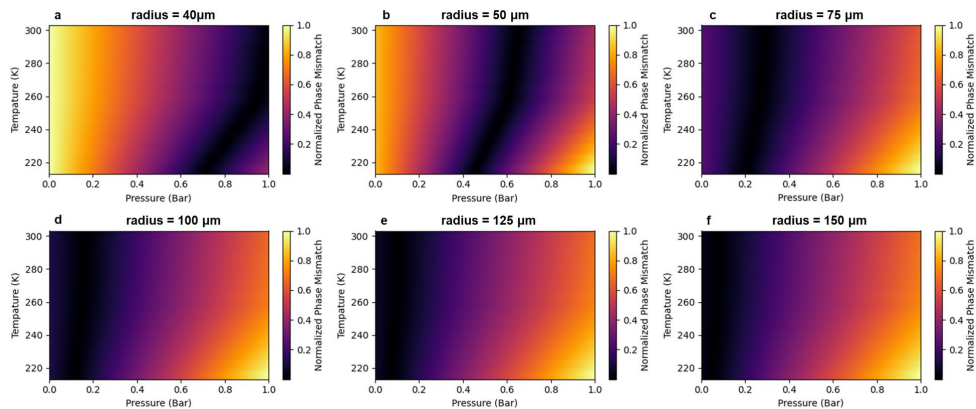

Figure S1. Graphical representation of phase mismatch variations under different pressures and temperatures using Ar.

**S-II: Linear spectral phase transfer with different pump's energy**

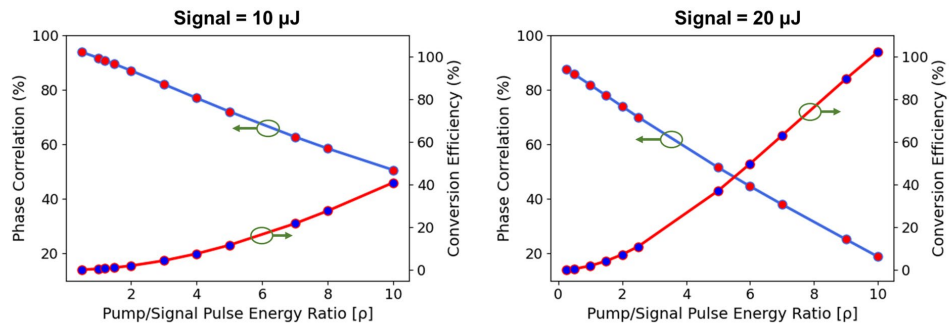

Figure S2. Illustration of the trade-off curve demonstrating the relationship between efficiency (red line with blue dots), defined as the ratio of idler pulse energy to signal pulse energy, and the quality of phase (blue line with red dots), characterized by linearity within the idler pulse. The curve is analyzed under two conditions: (a) with a signal pulse energy of 10 µJ, and (b) with a signal pulse energy of 20 µJ. The large conversion efficiency and the best linear phase transfer cannot be achieved at the same time.

**S-III: Phase transfer with different pump's FWHM**

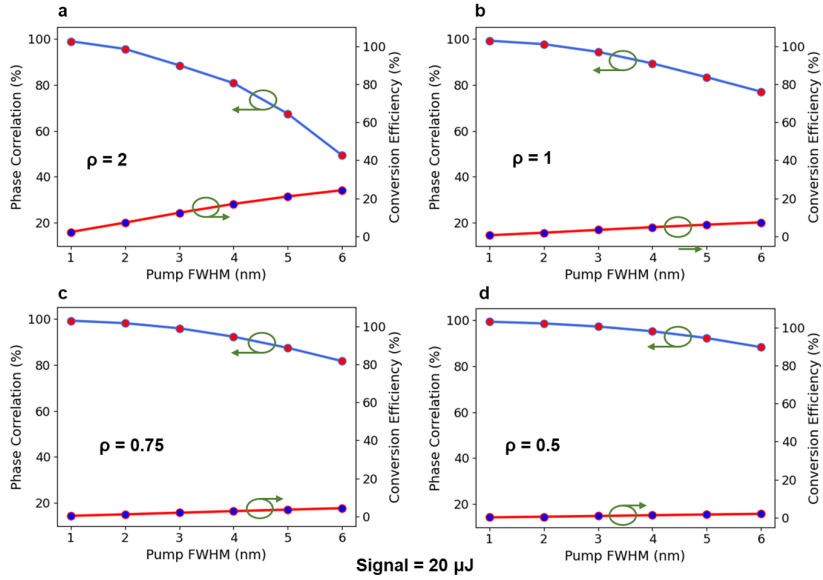

Figure S3. The interplay between efficiency, measured as the $\Delta\lambda_p$ (FWHM) of the pump spectrum, and the phase quality, represented by the linearity of the idler pulse with $\rho = 2$, $\rho = 1$, $\rho = 0.75$ and $\rho = 0.5$. The signal pulse energy is fixed at 20 μJ.

## S-IV: π-step phase transfer

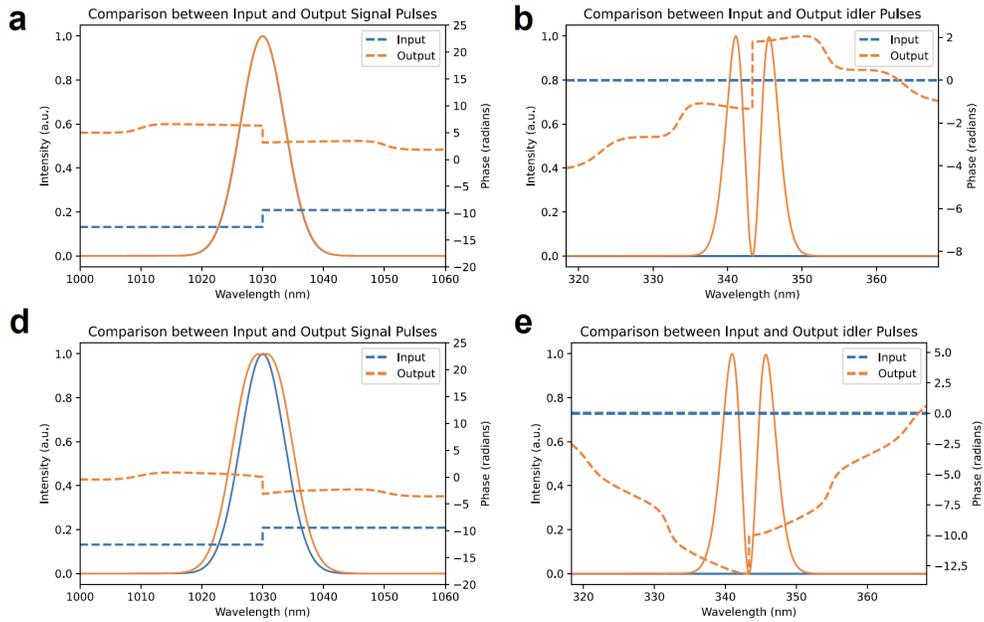

Figure S4. π-step phase transfer. (a-b) focused on better phase transfer from the signal (a) to idler (b) with $\rho = 1$. (c-d) focused on higher conversion efficiency signal (a) to idler (b) with

$\rho = 10$. The signal pulse energy is consistently set at 10 µJ. Due to the abrupt change in the spectral phase at the jumping point, the π-step behavior can still be transferred with low quality.